\documentclass[onecolumn,aps,floatfix,superscriptaddress]{revtex4-2}

\usepackage{amsmath}
\usepackage{graphicx}

\newcommand{\ti}{ \tilde }
\newcommand{\ep}{ \epsilon }
\newcommand{\pa}{ \partial }
\newcommand{\hb}{ \hbar }
\newcommand{\si}{ \sigma }

\newcommand{\ga}{ \gamma }

\newcommand{\cl}{ \mbox{cl} }

\newcommand{\im}{ \mbox{Im} }

\newcommand{\trho}{ \tilde{\rho} }

\newcommand{\hbt}{ \tilde{\hbar} }

\newcommand{\tiW}{ \ti{W} }

\begin{document}

%%%FIGURE SETUP - please do not change any commands within this section%%%

\title{Quantum-to-classical transition and H-theorem in surface diffusion} 

\author{E. E. Torres-Miyares}
\email{elena.torres@iff.csic.es}
\affiliation{Instituto de F\'isica Fundamental, Consejo Superior de Investigaciones Cient\'ificas, Serrano 123, 28006 Madrid, Spain}

\author{S. Miret-Art\'es}
\email{s.miret@iff.csic.es}
\affiliation{Instituto de F\'isica Fundamental, Consejo Superior de Investigaciones Cient\'ificas, Serrano 123, 28006 Madrid, Spain}

\begin{abstract}	
		
In this work, surface diffusion is studied with a different perspective by showing how the corresponding open dynamics is transformed when passing, in a continuous and smooth way, from a pure quantum regime to a full classical regime; the so-called quantum-to-classical transition. This continuous process is carried out from the Liouville-von Neumann equation by scaling Planck's constant. For this goal, the Brownian motion of an adsorbate on a flat surface is analyzed in order to show how this transition takes place. In particular, this open dynamics is studied from the master equation for the reduced density matrix within the Caldeira-Leggett formalism; in particular, the two extreme time behaviors, the ballistic and diffusive motions. It is also shown that the origin of the ballistic motion is different for the quantum and classical regimes. In this scenario, the corresponding Gaussian function for the intermediate scattering function is governed by the thermal velocity in the classical regime versus the initial spreading velocity of the wave packet for the quantum regime, leading to speak of classical and quantum ideal gas, respectively. Finally, in the diffusive regime, and starting from the Chudley-Elliott model, the quantum-to-classical transition is also discussed in terms of the well-known H-function for three surface temperatures in the diffusion of H and D on a Pt(111) surface. The main goal in this analysis is if one can discriminate the irreversibility coming from tunneling and thermal activation diffusion. \\%\\%The abstrast goes here instead of the text "The abstract should be..."
\end{abstract}
	
\maketitle

%%%FOOTNOTES%%%
\footnotetext{\textit{Instituto de F\'isica Fundamental, Consejo Superior de Investigaciones Cient\'ificas, Serrano 123, 28006 Madrid, Spain; E-mail:elena.torres@iff.csic.es; s.miret@iff.csic.es}}

%%%END OF FOOTNOTES%%%

%%%MAIN TEXT%%%%
\section{Introduction}

Surface diffusion is a crucial step in most elementary dynamical process occurring on surfaces. 
This diffusion process is, in general, analyzed by several experimental techniques where usually a probe
particle is interacting weakly with the system of interest coupled to a reservoir (or thermal bath) and measuring its linear response. The original theory dates back from van Hove when dealing with
neutron scattering by crystal and liquids. \cite{vanHove,Vineyard,Lovesey} 
The nature of particles (photons, neutrons, electrons or atoms) probing
systems formed by moving and interacting particles is largely 
irrelevant when the Born approximation is assumed, reducing 
this scattering event to a typical statistical mechanics problem. \cite{Lovesey,mcquarrie1976statistical} The response of the system is linear and determined by the spectrum of the spontaneous fluctuations of the reservoir 
according to the very well-known fluctuation-dissipation theorem. \cite{Lovesey} 

In most surface experimental techniques, the main observable magnitudes or response functions are the so-called dynamical
structure factor (DSF) and the intermediate scattering function (ISF). Quasi-elastic He atom scattering (QHAS), neutron scattering (QENS) together with their spin-echo (SE) techniques for Helium atoms (HeSE) and neutrons (NSE) have been widely developed. \cite{Hofman1996,Jardine2009a,Jardine2009b}
For fast diffusion motions of adsorbates, He atoms considered as nondestructive probe particles are generally used. These time of flight techniques are sensitive 
to surface processes on the length and time scales on which 
single atoms diffusion occurs (length and time scales covering 
10$^{-10}$- 10$^{-8}$ and 10$^{-12}$- 10$^{-8}$ seconds, respectively). 
Information provided by the experiment together with a 
theoretical support or theory behind can allow us to better understand the adsorbate dynamics by extracting 
valuable information for molecular interactions (adsorbate-substrate and adsorbate-adsorbate interactions) as well as
friction coefficients and surface corrugation from stochastic processes within
the general theory of quantum and classical open systems.
 
In surface diffusion, the so-called differential reflection
coefficient $\mathcal{R}$ (or the probability that the 
probing particles are scattered when interacting 
with adsorbates), forming a certain solid angle $\Omega$ with 
an energy exchange $\hbar \omega$ and a  momentum transfer parallel to surface, $\textbf{K}$, is proportional to the DSF, $S(\textbf{K},\omega)$, or response function according to
\begin{equation}\label{R}
	\frac{d \mathcal{R}^2 (\textbf{K},\omega)}{d \Omega \, \, d (\hbar \omega)}  = n_d \, F^2 \, S(\textbf{K},\omega),
\end{equation}
where $n_d$ is the diffusing particle concentration on the 
surface, $F$ is the atomic form factor which depends on 
the mutual interaction between probe particles and $N$ adsorbates. In this context, capital letters are often used to design 
variables on the surface. The DSF also gives the spectrum of spontaneous 
fluctuations.

Furthermore, another response function 
directly measured by HeSE experiments is the ISF, 
$I(\textbf{K},t)$, which is also obtained through the DSF by 
a frequency Fourier transform as follows
\begin{equation}\label{ISF1}            
I(\textbf{K},t)  =  \frac{1}{2 \pi} \int d \omega \, e^{ i \omega t} \, S(\textbf{K},\omega) = \frac{1}{N} \langle \rho_\textbf{K} \rho_\textbf{-K}(t) \rangle  , 
\end{equation}
where the particle density operator in the reciprocal space is 
$\rho_\textbf{K}(t) = \sum_{j=1}^N \exp(- i \textbf{K} . \textbf{R}_j(t))$; $\textbf{R}_j(t)$ being the position operator 
of the adsorbates as a function of time.
Following van Hove, within the Born approximation, $S$ and $I$ can 
be expressed in terms of the generalized pair-distribution function
$G(\textbf{R},t)$ (also known as the van Hove space-time correlation
function). This function gives the probability that given 
a particle at the origin at time $t=0$, any particle, including 
the same one, can be found at the position $\textbf{R}$ and at 
time $t$. This function can also be extracted from the inverse space Fourier transform of the ISF as
\begin{equation}\label{G}            
G(\textbf{R},t)  = \int d \textbf{K} \, e^{- i \textbf{K} . \textbf{R}} \, I(\textbf{K},t). 
\end{equation}
Thus, from Eqs. (\ref{ISF1}) and (\ref{G}), $S(\textbf{K},\omega)$ is expressed as a density-density correlation function,

A fundamental aspect of the linear response theory is the quantum-to-classical transition.\cite{mcquarrie76a,Zwanzig2001} There are several ways to study this transition in the literature within the theory of open quantum systems.
%the most well-known one being the WKB formalism where a Taylor series
%expansion in powers of Planck's constant is considered. In particular, 
%the classical Hamilton-Jacobi equation for the classical action is obtained at zero order and a hierarchy of differential equations for the action at different orders of the expansion is finally obtained. Moreover, it is also well established the classical limit of the Heisenberg commutator to be the classical Poisson bracket through the Weyl quantization rule.
For a closed quantum system, apart from the well-known WKB formalism, a different approach maybe less known in the literature is due to 
Schiller \cite{Schiller1, Schiller2, Schiller3} and Rosen \cite{Rosen1, Rosen2} who proposed and discussed what is known as the classical wave function ruled by a nonlinear differential wave equation. The nonlinearity is precisely due to the so-called quantum potential from Bohmian mechanics. \cite{Holland,Salva-b1,Salva-b2} 
As mentioned by Holland, a nonlinear differential equation may be a limiting case of a linear equation, the time-dependent Schr\"odinger equation. Afterwards Richardson {\it et al.}
\cite{Richardson2014} introduced a nonlinear quantum-to-classical transition  differential wave equation by means of a transition parameter $\epsilon$, with $\epsilon \in [0,1]$, going continuously and smoothly from the quantum regime, the standard Schr\"odinger wave equation ($\epsilon =1$), to the classical regime ($\epsilon=0$) in the sense of Schiller. 
They also proved the equivalence of this nonlinear differential equation with a linear one.
The interesting point is that quantum features are observed in all in-between dynamical regimes but they are fading gradually and continuously with decreasing $\epsilon$ in a nonlinear way (going as $\sqrt{\epsilon}$). This decoherence scheme is different from the well-known decoherence process which involves the time evolution of the system in consideration in order to reach  classical features.  Interesting and simple applications 
of this transition differential wave equation are found with Gaussian wave packets. \cite{Chou1,Chou2} Very recently, the so-called bouncing ball problem has also been solved within this framework.\cite{Mousavi2024} 
The same procedure has been used for open quantum systems by means of wave functions.\cite{Nassar,Mousavi2022a}
One then speaks about the scaled Schr\"odinger differential equation or, in general, on the {\it scaled quantum theory}.\cite{Mousavi2022a,Mousavi2024}

On the other hand, in non-equilibrium statistical mechanics, it is well-known that the quantum analog of the classical phase space distribution is 
the quantum mechanical density matrix. In classical theory, one needs to know the time-dependent distribution function to calculate time-dependent averages of any dynamical variable or correlation functions. The corresponding time evolution is governed by the classical Liouville equation. In the 
quantum domain, the quantum Liouville-von Neumann equation, which is also associated with a Hamiltonian, gives us the time evolution of the density matrix. Both equations have the same formal structure but the Liouville operator is obviously written differently. \cite{Zwanzig2001} In general, the equations of motion satisfied  by classical and quantum distribution functions enters within the domain of the so-called transport or kinetic theory, starting from the corresponding Liouville equations.\cite{Liboff} 
Phase and position time correlation functions are also governed by kinetic equations. In this context, the Wigner distribution and the Wigner-Moyal equation play a central role in any quantum-to-classical correspondence.
Recently, following Richardson's procedure, we have shown
that both worlds are connected in a continuous and smooth way for all the dynamical regimes in between these two extreme cases by using  the  scaled Liouville-von Neumann equation.\cite{Mousavi2023} This could be the starting point to analyze problems in non-equilibrium statistical mechanics or, in general, in open quantum systems.

In our context, surface diffusion sampled by He atoms, the van Hove pair distribution function $G(\textbf{R},t)$ has been replaced by the reduced density matrix $\rho(\textbf{R},t)$.\cite{Torres2022,Torres2023} Two theoretical formalisms have been used, the Caldeira-Leggett (CL) master equation \cite{Caldeira1983} and the Lindblad equation. \cite{Lindblad}
%In this work, we are going to focus on the CL formalism \cite{CaLe-PA-1983,Caldeira-2014}. 
The CL formalism is based on
the reduced density matrix once one carries out the integration over the environmental degrees of freedom. As is well-known, the diagonal matrix
elements give us probabilities and off-diagonal matrix elements, coherences. In the decoherence process, these off-diagonal elements 
go to zero with time more or less rapidly depending on the parameters characterizing the environment; usually, the friction coefficient and the surface temperature. Most of
studies involving quantum decoherence are being carried out in the configuration space and very few in the momentum space.\cite{Mousavi2021}
One of the main purposes then of this work is to analyze the quantum-to-classical transition of the response functions presented above within the scaled theory.

The second important aspect analyzed here is the implications of this transition in the so-called ballistic motion (when time is much less that the inverse of the friction coefficient). Thus, the origin of the $t^2$-behavior is different in the quantum and classical regime. Even more, this analysis shows that, due to the fact the ballistic motion is frictionless, one can speak of classical and quantum ideal gas and, therefore, there is no need to invoke an effective mass of the adsorbate in order to better understand this ballistic motion.

The third fundamental aspect dealing with 
this transition is when analyzing surface diffusion by tunneling (with light adsorbates); in particular, the coexistence of activated and tunneling diffusion. The parameter controlling this transition is surface temperature. 
The knowledge of activated and tunneling rates as a function of the 
surface temperature in an Arrhenius-like plot provide very 
valuable information about surface structure and adsorbate-substrate interaction. In particular, the physical parameters which can be extracted are at least the 
friction coefficient (if Ohmic friction is assumed), barrier heights, well and barrier frequencies and anharmonicity if parabolic barriers are not good enough 
to describe this open dynamics. Within Kramers' formalism, \cite{Eli2023} analytical expressions are obtained when several approximations are assumed and this is the starting point for a  fitting procedure in order to extract those physical parameters. As said above, this transition can be carried out now by varying the surface temperature covering thermal activation (high temperatures) and tunneling above and below the so-called crossover temperature which provides us the region where quantum effects start to be determinant. The diffusion of the H and D adsorbates on a Pt(111) surface is paradigmatic in this regard, speaking about incoherent tunneling if the surface temperature is not too low (at very low temperatures, tunneling is coherent). Anomalous friction coefficients are reported in the literature. \cite{jardine2010determination,Ruth2013} Recently, a new study on incoherent tunneling surface diffusion for these systems has been published by solving the Lindblad master equation following the so-called stochastic wave function (SWF) method \cite{Elena2025} due to the fact that Kramers' theory is not able to be applied so far because of the breakdown of some approximations used in this theory. 

In any diffusion jump model, a Pauli (gain-loss) master equation is generally assumed. Master equations describe the dynamics of transitions between states. Like a kinetic equation, rate constants are replaced by transition rates and concentrations by state probabilities. Master equation are inherently linear. One way of solving this equation and widely used in this context is due to Chudley-Elliott (CE).\cite{Chudley} The main drawback of the CE model is that the transition rates are fitting parameters  and, therefore,  not known {\it a priori}. When Kramers turnover theory is applicable can also provide a prediction for all jump probabilities.\cite{Eli2023} The ultimate goal, in each case, is to know the total hopping rates versus the inverse of the surface temperature in an Arrhenius plot as well as to elucidate if the diffusion is thermal activated or proceeds via tunneling. Recently, we have shown that the ISF can be seen as a characteristic function from probability theory within the general formalism of the master equation.\cite{Torres2025a}
One aspect, not considered in this context so far, is that the Pauli master equation is related to the well-known H-theorem of non-equilibrium statistical mechanics.\cite{Liboff} When speaking about the H-theorem, one thinks in terms of irreversibility and entropy; in other words, it describes how some functions (entropy, S, and H functions) evolve with time towards the thermal equilibrium. In particular, we are interested if one can discriminate the irreversibility coming from tunneling and thermal activation diffusion. For this goal, we have analyzed the diffusion of H,D on a Pt(111) surface where the quantum character of the particles as well as the surface temperature play an important role in the evolution of the H function and entropy; in particular, in the quantum-to-classical transition and close to the cross-over temperature.

This paper is then organized as follows. In Section 2, the scaled Liouville-von Neumann equation is presented and discussed. In Section 3, the scaled theory is applied to the Brownian motion on a flat surface in order to obtain closed expressions for a better discussion. In particular, the flux of a Brownian particle and the ISF and DSF response functions are analyzed for the two typical extreme time behaviors, short time or ballistic motion and long time or diffusive motion. In Section 4, the general diffusion process is analyzed under the quantum-to-classical transition perspective by considering the thermal activation and incoherent tunneling surface diffusion in terms of the H-theorem. Finally, in Section 5, some conclusions are presented.

\section{The scaled Liouville-von Neumann equation and its Wigner representation} 

In quantum mechanics, the density operator $\hat \rho$ is governed by the the so-called Liouville-von Neumann equation of motion (or simply von Neumann equation), which is written as
\begin{eqnarray} \label{eq: Neumann}
i \hb \frac{\pa \hat{\rho}}{\pa t} &=& [\hat{H}, \hat{\rho}]  ,
\end{eqnarray}
where $ \hat{H} $ is the Hamiltonian operator of the system and $[\cdot, \cdot]$ gives the commutator 
of two operators. For simplicity, let us consider a one-dimensional problem and only one particle. This 
Hamiltonian is expressed as the sum of the kinetic energy operator for a particle of mass $m$ and an external potential $V( \hat{x} )$
as follows
\begin{eqnarray}
\hat{H} &=& \frac{ \hat{p}^2 }{2m} + V( \hat{x} )   .
\end{eqnarray}
Eq.  (\ref{eq: Neumann}) in the position representation reads thus as
\begin{eqnarray} \label{eq: vn_eq}
	i\hb \frac{\pa}{\pa t} \rho(x, x', t) &=& - \frac{\hb^2}{2m} 
	\left( \frac{\pa^2}{ \pa x^2 } - \frac{\pa^2}{ \pa x'^2 }   \right) \rho(x, x', t) \nonumber \\
	&+& \big(V(x) - V(x')\big) \rho(x, x', t)   . 
\end{eqnarray}
As mentioned above, diagonal elements of the density matrix give probabilities while the non-diagonal elements represent coherences, which provide the correlation of  the particle at two different positions $x$ and $x'$ at time $t$. With time, this correlation or coherence decays obviously to zero. 
From Eq. (\ref{eq: vn_eq}), one can easily reach the continuity equation written as
\begin{eqnarray}\label{ceq}
\frac{\partial \rho (x,x',t)}{\partial t} \big|_{x=x'} + \frac{\partial j(x,t)}{\partial x}&=& 0   ,
\end{eqnarray}
with the probability current density given by  
\begin{eqnarray}\label{j}
j(x, t) &=& \frac{\hb}{m} \im \left\{ \frac{\pa}{\pa x} \rho(x, x', t)\bigg|_{x=x'} \right \}   .
\end{eqnarray} 

Recently, following the same spirit of Schiller and Rosen \cite{Schiller1,Rosen2} 
who proposed a classical wave equation, we have proposed a von 
Neumann equation but for the classical density matrix expressed as 
\cite{Mousavi2023}
\begin{eqnarray} \label{eq: cl-vN}
& & i\hb \frac{\pa}{\pa t} \, \rho_{\cl}(x, x', t) = - \frac{\hb^2}{2m} 
	\left( \frac{\pa^2}{ \pa x^2 } - \frac{\pa^2}{ \pa x'^2 }   \right) \rho_{\cl}(x, x', t) \nonumber \\
	&+& (V(x) - V(x')) \rho_{\cl}(x, x', t)
	\nonumber \\
	&+& \frac{\hb^2}{2m} \left [ \frac{1}{ \big| \rho_{\cl}(x, x', t) \big|} \left( \frac{\pa^2}{ \pa x^2 } - \frac{\pa^2}{ \pa x'^2 }   \right) \big| \rho_{\cl}(x, x', t) \big| \right ] \rho_{\cl}(x, x', t)  , \nonumber \\
\end{eqnarray}
where {\it cl} refers to {\it classical} and $ | \rho_{\cl}(x, x', t) | $ means the modulus of $ \rho_{\cl}(x, x', t) $.
This classical von Neumann equation will serve us as a starting point 
to describe the quantum-to-classical continuous transition in the same 
way  a {\it non-linear} transition wave equation  was proposed within 
the Schr\"{o}dinger framework, previously. \cite{Richardson2014} 
If one now defines a {\it scaled Planck constant} as
\begin{eqnarray} \label{eq: hbt}
	\hbt &=& \sqrt{\ep}~ \hb  ,
\end{eqnarray}
which contains a {\it transition parameter} $\epsilon$ going from $\epsilon = 1$ (quantum regime) to $\epsilon = 0$  (classical regime). Notice that the transition is non-linear with $\epsilon$, it is slower than in the classical standard limit. The corresponding non-linear von Neumann equation is rewritten as 

%\cite{MoMi-Sym-2023}
%
\begin{eqnarray} \label{eq: tran-vN}
& & i\hb \frac{\pa}{\pa t} \, \rho_{\epsilon}(x, x', t) = - \frac{\hb^2}{2m} \left( \frac{\pa^2}{ \pa x^2 } - \frac{\pa^2}{ \pa x'^2 }   \right) \rho_{\epsilon}(x, x', t) \nonumber \\
&+& (V(x) - V(x')) \rho_{\epsilon}(x, x', t)  \nonumber \\
&+& (1-\epsilon) \frac{\hb^2}{2m} \left [ \frac{1}{ | \rho_{\epsilon}(x, x', t) |} \left( \frac{\pa^2}{ \pa x^2 } - \frac{\pa^2}{ \pa x'^2 }   \right) | \rho_{\epsilon}(x, x', t) |  \right ] \rho_{\epsilon}(x, x', t)    ,  \nonumber \\
\end{eqnarray}
or as
\begin{eqnarray} \label{eq: scaled-vN}
i\hbt \frac{\pa}{\pa t} \ti{\rho}(x, x', t) &=& - \frac{\hbt^2}{2m} 
\left( \frac{\pa^2}{ \pa x^2 } - \frac{\pa^2}{ \pa x'^2 }   \right) \ti{\rho}(x, x', t) \nonumber \\
&+& (V(x) - V(x')) \ti{\rho}(x, x', t)   .
\end{eqnarray}
This is the so-called {\it scaled von Neumann equation}.
The form of this scaled equation is exactly the same as that of the 
linear von Neumann equation. 
The only changes are that $\hb$ and $\rho$ have been replaced by the corresponding scaled 
quantities, $ \hbt $ and $ \ti{\rho} $. Interestingly enough, all the in-between dynamical regimes covering the interval $[0,1]$ display quantum features except for values of $\ep$ approaching to zero.
Thus, all requirements for solutions of  the standard von Neumann equation  must also be fulfilled. 
The scaled von Neumann equation (\ref{eq: scaled-vN}) can be recast in the form
\begin{eqnarray} \label{eq: scaled-vN2}
	i\hbt \frac{\pa}{\pa t} \hat{\trho} &=& [\hat{\ti{H}}, \hat{\trho}]  ,
\end{eqnarray}
where the scaled Hamiltonian operator in the position representation is expressed as 
\begin{eqnarray} \label{eq: scaled-Ham}
	\ti{H} &=& -\frac{\hbt^2}{2m} \frac{\pa^2}{\pa x^2} + V(x)  ,  
\end{eqnarray}
where the scaled Planck constant appears only in the kinetic operator. Furthermore, the structure of the scaled continuity equation is then the same as written in Eq.(\ref{ceq}), with the scaled probability density current given by
\begin{eqnarray} \label{eq: scaled-pcd}
\ti{j}(x, t) &=& \frac{\hbt}{m} \im \left\{ \frac{\pa}{\pa x} \ti{\rho}(x, x', t)\bigg|_{x=x'} \right \} .
\end{eqnarray} 
%

%Furthermore, in general, the time derivative of an arbitrary time-independent observable $ \hat{A} $ is calculated from
%
%\begin{equation} \label{eq: dAdt}
%\frac{d}{dt} \wti{ \la \hat{A} \ra} =  \left ( \frac{\pa \hat{\trho} }{\pa t} \hat{\tilde A} \right ) 
%= \frac{1}{i \hbt} \wti{ \la [\hat{A}, \hat{\ti{H}}] \ra } % . 
%\end{equation}
%

By changing now to the center of mass and relative coordinates defined as $R = (x+x')/2$ and $r = x - x'$ and similarly for the momentum $ u = (p+q)/2 $ and $ v = p-q $ respectively, 
the Fourier transform of the scaled density matrix $ \ti \rho(R,r,t) $ with respect 
to the relative coordinate $ r $ can be expressed as
\begin{eqnarray}
	\tiW(R,u,t) &=& \frac{1}{2 \pi \hbt} \int dr ~ e^{-i u r / \hbt} \trho(R,r,t)  , \label{eq: F-rho} %\\
%	&=& \frac{1}{2 \pi \hbt} \int dr ~ e^{-i u r / \hbt} \tpsi(R+r/2, t) \tpsi^*(R-r/2, t)  , \nonumber 
	%\\&=& \frac{1}{\pi \hbt} \int dr' ~ e^{i u r' / \hbt} \tpsi(R-r', t) \tpsi^*(R+r', t) \nonumber
\end{eqnarray}
which is just the {\it scaled} Wigner distribution function. As is well-known, this distribution serves as a bridge between quantum and classical kinetic theory. \cite{Liboff} 
%
%\begin{eqnarray}
%\ti{\rho}(R, r, t) 
%&=& \frac{1}{2 \pi \hbt} \int du dv~ \ti \phi(u+v/2,t) \ti \phi^* (u-v/2,t) e^{i (R v + r u) / \hbt } \nonumber \\
%	& = & \int du ~ \tiW(R,u,t) e^{i u r / \hbt}   \label{eq: rho-F}  ,
%\end{eqnarray}
%
%where we have defined 
%
%\begin{eqnarray} \label{F}
%\tiF(R,u,t) &=& \frac{1}{2 \pi \hbt} \int dv~ \ti \phi(u+v/2,t) \ti \phi^* (u-v/2,t) e^{i R v / \hbt }.
%\end{eqnarray}
%
%where $ \tiW(R,u,t) $ is the Fourier transform of the scaled density matrix $ \trho(R,r,t) $ with respect 
%to the relative coordinate $ r $ or equivalently
%This can be explicitly seen by changing the relative variable $ r \to - r/2 $. 
The time evolution of the scaled Wigner distribution function $ \tiW(R,u,t) $ (scaled kinetic equation) can be found from 
Eqs. (\ref{eq: F-rho}) and (\ref{eq: scaled-vN}), written in the coordinates $x$ and $p$, according to
\begin{equation}\label{eq: dWdt}
\frac{\partial \ti W(x, p, t)}{\partial t} + \frac{p}{m} \frac{\partial \ti W(x,p,t)}{\partial x}
	 = \int \ti{K}(x,p-q) \ti W(x,q,t) dq	,
\end{equation}	
where the kernel $	\ti{K}(R, q,t) $ is given by
\begin{eqnarray} \label{eq: J-kernel}
	\ti{K}(R, q,t) &=& \frac{1}{2 \pi \hbt} \int dr ~ e^{i q r / \hbt} \frac{ U(R/2+r) - U(R/2-r) }{i\hbt} .
\end{eqnarray}
%
%\begin{eqnarray} \label{eq: dFdt}
%	\frac{\pa}{\pa t}\tiW(R, u, t) &=& \frac{1}{2 \pi \hbt} \int dr ~ e^{-i u r / \hbt} \frac{\pa}{\pa t} \trho(R,r,t)  \nonumber \\  
%	&=& \frac{1}{2 \pi \hbt} \int dr ~ e^{-i u r / \hbt} \left( \frac{i\hbt}{m} \frac{\pa^2}{\pa r \pa R}
%	+ \frac{ U(R/2+r) - U(R/2-r) }{i\hbt} \right) \trho(R, r, t) \nonumber \\
%	&=& - \frac{u}{m}  \frac{\pa}{\pa R} \tiW(R, u, t) 
%	+ \int du' \ti{K}(R, u'-u, t) \tiW(R, u',t)    ,
%\end{eqnarray}
%
The $\ti W(x,p,t) $ distribution function satisfies the Liouville equation in the classical domain \cite{Liboff} as well as the corresponding scaled distribution. Furthermore,
the so-called Wigner-Moyal equation in quantum theory is the analog of the particle Liouville equation in kinetic theory. In the scaled
framework,  we have shown that this Wigner-Moyal equation 
can be written as \cite{Mousavi2023}

%In particular, the time derivative of the corresponding characteristic function assumed to satisfy a quantum Liouville equation 
%
%\begin{equation}
%i \ti \hbar \frac{\partial \hat M}{\partial t} = [\ti H, \hat M]	,
%\end{equation}
% 
%with $\ti H = - \frac{- \ti \hbar^2}{2 m} \frac{\partial^2}{\partial x^2} + U(x)$. 
%
%
\begin{eqnarray} \label{eq: F-evol}
\frac{\pa}{\pa t} \tiW(x, p, t) 
    &=& \frac{2}{\hbt} 
	 \sin \left[ \frac{\hbt}{2} \left( \frac{\pa}{\pa p_{\tiW}} \frac{\pa}{\pa x_H} - \right . \right .\nonumber \\
	&-& \left . \left . \frac{\pa}{\pa p_H } \frac{\pa}{\pa x_{\tiW}}
	\right)  \right]  H(x, p) \tiW(x, p, t)   , 
\end{eqnarray}
where $ H(x, p) $ is the classical Hamiltonian, and $ \pa/ \pa x_{\tiW} $ and $ \pa/ \pa p_{\tiW} $ operates only 
on $  \tiW $. In the classical limit $ \ep \to 0 $, and keeping the leading term in the expansion of the sine-function, this equation reduces to the well-known classical Liouville equation for the 
phase space distribution function,
\begin{eqnarray} \label{eq: Lio}
	\frac{\pa}{\pa t} W_{\cl}(p, q, t) &=& \{ W_{\cl}, H \}_{PB}   ,
\end{eqnarray}
where $ \{ \cdot, \cdot \}_{PB} $ stands for the Poisson bracket.
Thus, in this scaled non-equilibrium statistical mechanics, all the in-between dynamical regimes are issued from a continuous and smooth way  when going from the quantum to the classical regime. 
Thus, an alternative and equivalent way to pass from a quantum to classical Liouville equation in a continuous way, different from what shown in standard books dealing on quantum kinetic theory, \cite{Liboff} is proposed. 
%In standard textbooks of nonlinear statistical mechanics, \cite{mcquarrie76a,Zwanzig} it is established without demonstration a correspondence between classical and quantum expressions; for example, the quantum commutator by the Poisson bracket, the trace operation by a classical average in the phase-space, etc. In very few studies, one can find the alluded correspondence from Heisenberg matrix formalism of quantum mechanics but for conservative systems. 

\section{The scaled Brownian motion on a flat surface}
\subsection{The quantum and classical regimes}

Due to the fact that the quantum Brownian motion of a particle is the paradigm of an open quantum system, \cite{feynman1964brownian} the corresponding free motion is the easiest and 
simplest application one can devise. A complete mathematical model for the system-plus-environment dynamics is usually quite complicated. The environment is in general well represented by a reservoir with an infinite number of degrees of freedom, a bosonic bath consisting of an infinite number of quantum oscillators in thermal equilibrium, and affects the system of interest through a position-position coupling.\cite{breuer2002theory} This dynamics can be described by an equation of motion for the corresponding density matrix. By assuming short environmental correlation times, the memory effects can be safely neglected in the reduced dynamics and a Markovian quantum master equation can be easily derived which is known as the CL master equation. \cite{Caldeira} This formalism gives the evolution of the reduced density matrix in the coordinate representation by tracing out the environment degrees of freedom and contains both frictional and thermal effects due to the environment. \cite{Caldeira,breuer2002theory} 
The corresponding Markovian equation in the coordinate representation for one dimension, at the high temperature limit and for a particle of mass $m$, is written as \cite{Caldeira}
\begin{eqnarray} \label{eq: CL eq}
\frac{\pa \rho(x, x', t)}{\pa t} &=& \left[ - \frac{\hb}{2mi} \left( \frac{\pa^2}{\pa x^2} - \frac{\pa^2}{\pa x'^2} \right) - \ga (x-x') \left( \frac{\pa}{\pa x} - \frac{\pa}{\pa x'} \right) \right. \nonumber \\
&+& \frac{ V(x) - V(x') }{ i\hb } - \left.  \frac{D}{\hb^2} (x-x')^2 \right] \rho(x, x', t)  
\end{eqnarray}
where $V$ is the external interaction potential, $\gamma$ the friction coefficient (Ohmic friction) and 
\begin{eqnarray} \label{eq: D}
D &=& 2 m \ga k_B T  
\end{eqnarray}
plays the role of the diffusion coefficient; $k_B$ and $T$ being Boltzmann’s constant and the environment temperature, respectively.

We have stressed the fact that the scaled time evolution for the density matrix has the same formal expression as the standard one. In this sense, one could extend in a quite straightforward way the same studies carried out in the standard quantum mechanics to the scaled quantum theory (see, for example, Ref. \cite{Mousavi2024}).  The scaled CL master equation for the reduced density matrix in the position representation for one dimension, at the high temperature limit, takes thus the form of a Markovian equation
\begin{eqnarray} \label{eq: sCL eq}
	\frac{\pa \ti{\rho}(x, x', t)}{\pa t} &=& \left[ - \frac{\hbt}{2\,m\,i} \left( \frac{\pa^2}{\pa x^2} - \frac{\pa^2}{\pa x'^2} \right) - \ga\, (x-x') \left( \frac{\pa}{\pa x} - \frac{\pa}{\pa x'} \right)
	+  \right. \nonumber \\
	&+& \left. \frac{ V(x) - V(x') }{ i\hbt } -  \frac{D}{\hbt^2} (x-x')^2 \right] \ti{\rho}(x, x', t),  
\end{eqnarray}
where we have just replaced, as mentioned before, $\hb$ and $\rho$ by  $ \hbt $ and $ \ti{\rho} $, respectively.
%Obviously, the same discussion and expressions are obtained for the scaled theory by considering the replacements above mentioned. 
Then, one has the scaled continuity equation written as
\begin{eqnarray} \label{eq: scon_CL}
\frac{\pa \ti \rho(x, t)}{\pa t} + \frac{\pa \ti j(x, t)}{\pa x}  &=& 0 .
\end{eqnarray}
%
%and the flux as 
%
%\begin{eqnarray} \label{flux}
%\ti j(x,t) = \ti u(x,t) \, \ti \rho(x, t) ,
%\end{eqnarray}
%
%where $\ti u(x,t)$ is the scaled generalized velocity.

%\section{Dynamics of a Gaussian wave packet}

For flat surfaces ($V=0$), analytical solutions are easily obtained by considering, for example, an initial Gaussian wave packet {\it ansatz} \cite{schlosshauer2007decoherence} 

\begin{eqnarray} \label{eq: wf0}
\ti \psi_0(x) &=& \frac{1}{(2\pi \si_0^2)^{1/4}} \exp \left[ - \frac{(x-x_0)^2}{4  \si_0^2} + i \frac{p_0}{\ti \hb} x \right],
\end{eqnarray}
$x_0$, $p_0$ and $\sigma_0$ being the initial values for the center, momentum and width, respectively. Eq. (\ref{eq: sCL eq}) is solved 
in three steps; first, by applying the technique of the Fourier transform with respect to the center of mass coordinate; second, solving the 
resulting equation and finally taking the inverse Fourier transform of  this solution in order to obtain the density matrix in the position representation.\cite{Mousavi-2022} In this way, one obtains
\begin{eqnarray}
\ti \rho(R, r, t) &=& \frac{1}{ \sqrt{2\pi} \ti \sigma_t } \exp\left[ \ti a_0(r, t) - \frac{ \big( R - \ti a_1(r, t) \big)^2 }{ 2 \ti \sigma_t^2} \right]   , \label{eq: denmat}  \\
\ti j(R, r, t) &=& -i\frac{\ti \hb}{m} \left( \frac{\pa \ti a_0}{\pa r} + \frac{ x- \ti a_1(0, t) }{\ti \sigma_t^2} \frac{\pa \ti a_1}{\pa r} \right) \ti \rho(R, r, t) , \label{eq: curmat}
\end{eqnarray}
for the non-diagonal elements of the density matrix (coherences) and the current density matrix with 
\begin{eqnarray}
\ti a_0(r, t)& =& - \left[ \frac{ e^{-4\ga t} }{ 8 \si_0^2 } + \frac{ 1 - e^{-4\ga t} }{4\ga} \frac{D}{\ti \hb^2} \right]r^2 + i  \frac{p_0}{\ti \hb} r ~e^{-2\ga t}   ,
\label{eq: a0}
\end{eqnarray}
and
\begin{eqnarray}
\ti a_1(r, t) &=& x_t  
+ i f(t) \left[ \frac{ \ti \hb }{ 4 m \si_0^2 } e^{-2\ga t}  
+ \frac{D}{m \ti \hb} f(t)  \right] r    ,
\label{eq: a1} 
\end{eqnarray}
and the time dependent width 
\begin{eqnarray}
\ti \sigma_t^2 &=& \si_0^2  + \frac{ \ti \hb^2 }{ 4 m^2 \si_0^2 }   f(t)^2
+ \frac{ 4\ga t + 4 e^{-2\ga t} - 3 - e^{-4\ga t} }{8 m^2 \ga^3 } ~ D  , \nonumber \\
&\equiv& \si^2_{0} - \frac{[x(0),x(t)]^2}{4\, \si_{0}^2} + \big\langle\{x(t)-x(0)\}^2\big\rangle  ,
\label{wt}
\end{eqnarray}
with
\begin{equation}\label{ft}
f(t) = \frac{1-e^{-2\ga t}}{2\ga} .
\end{equation}
Eq. (\ref{wt}) can also be found in the literature.
\cite{PhysRevA.64.032101,Ford_2003,Pe_ate_Rodr_guez_2012} 
%
%\begin{equation}\label{width}
%\ti \si_t^2 = \si^2_{0} - \frac{[x(0),x(t)]^2}{4\, \si_{0}^2} + \langle\{x(t)-x(0)\}^2\rangle_q  
%\end{equation}
% 
Three contributions are easily identified: the initial width, the commutator through 
$-[x(0),x(t)]^2/4 \sigma_0^2$ and the mean square displacement (MSD).

Thus, for a Brownian particle without external potential, the probability density and the probability current density, by imposing the condition $r=0$, are  
\begin{eqnarray}
\ti \rho(x, t)& =& \frac{1}{ \sqrt{2\pi} \ti \sigma_t } \exp\left[ - \frac{ ( x - x_t )^2 }{ 2 \ti \sigma_t^2} \right]  , \label{eq: probden}  \\
\ti j(x, t) &\equiv& \ti u(x,t) \, \ti \rho(x,t) = \left\{ \frac{p_0}{m} ~e^{-2\ga t} + \right. \nonumber \\
&+& \left. \frac{ x - x_t }{\ti \sigma_t^2} f(t)  \left[ \frac{ \ti \hb^2 }{ 4 m^2 \si_0^2 } e^{-2\ga t} + \frac{D}{m^2} f(t)  \right] \right\}  \ti \rho(x, t) , \nonumber \\
%&=& (\ti \dot x(t) + \ti u_D(x,t) \, \ti \rho(x,t)    ,
\label{eq: cur}
\end{eqnarray}
respectively. Equation (\ref{eq: probden}) then shows that the probability density has a Gaussian shape with a width $\ti \sigma_t$ according to Eq. (\ref{wt}) and a center moving along the {\it classical} trajectory $x(t)=x_t$ given by 
\begin{equation}\label{xt}
x_t = x_0 + v_0 f(t)     ,
\end{equation}
with $x(0)=x_0$. Usually, the adsorbate initial velocity is a thermal velocity expressed as $v_0 = \sqrt{2 k_B T/m}$.
For the probability flux given by Eq. (\ref{eq: cur}), the generalized velocity $\ti u$ is then expressed in terms of 
three contributions
\begin{equation}
\ti u(x,t) = v_0 ~e^{-2\ga t} + \frac{x - x_t }{\ti \sigma_t^2} f(t) \left(\ti v_s^2 ~e^{-2\ga t} + v_0^2 \ga \,f(t) \right)   .
\end{equation}
The first contribution is purely classical and shows how the initial thermal velocity is decreasing with time due to friction. The second term, $\ti v_s = \ti \hb / 2 m \si_0 $, is of quantum origin since it includes the initial spreading velocity of the wave packet. \cite{Salva-b2} And, finally, the last term depends also on the temperature due to the presence of the thermal velocity or equivalently on the diffusion coefficient; this contribution can be seen as a dispersive velocity.
%{\bf Note: maybe could be interesting to analyze the flux-flux autocorrelation function.}

When approaching the classical regime ($\epsilon \rightarrow 0$) \cite{chandrasekhar1943stochastic,mcquarrie1976statistical}, the Brownian particle can be 
described through a velocity or position stochastic process. In the first description, the Langevin equation 
and its equivalent Fokker-Planck (FP) equation for the probability distribution is written in terms 
of the velocity of the Brownian particle which is a stochastic variable (the so-called 
Ornstein-Uhlenbeck process). Due to the fact the density matrix we have used is written in the coordinate representation, the Brownian particle is chosen to be described by the position stochastic process. Thus, the position $x(t)$ can be written as
\begin{equation}
d x(t) = v(t) dt    ,
\end{equation}
$v(t)$ being the Brownian particle velocity; usually, it is assumed that $x(0) \neq 0$. The classical distribution function for the position of the Brownian particle, after the corresponding FP equation, which gives the probability to find a Brownian particle at position $x$ and at time $t$ when initially was at $x_0$ and $t=0$, is a Gaussian function given by 
\begin{equation}\label{crho}
\rho_{cl}(x, t) = \frac{1}{ \sqrt{2\pi \Big\langle\big(x(t)-x(0)\big)^2\Big\rangle_{cl}}} \exp\left[ - 
\frac{ ( x - x_t )^2 }{ 2 \Big\langle\big(x(t)-x(0)\big)^2\Big\rangle_{cl}} \right]    ,
\end{equation}
fulfilling the standard diffusion equation.
The mean particle position is expressed, in terms of the $f$-function, Eq.(\ref{ft}), as  
\begin{equation}
\langle x(t)\rangle = v_0 f(t)  ,
\end{equation}
and the classical MSD \cite{mcquarrie1976statistical} as
\begin{equation}\label{cwt}
\Big\langle\big(x(t)-x(0)\big)^2\Big\rangle_{cl} = v_0^2 f(t)^2 + \frac{ 4\ga t + 4 e^{-2\ga t} - 3 - e^{-4\ga t} }{8 m^2 \ga^3 } ~ D   .
\end{equation}
By comparing this classical MSD and that appearing in Eq. (\ref{wt}), apart from the initial width of the wave packet, one notes that the $t^2$-term is replaced by  the commutator of two time positions and the second term is however identical. 
Moreover, the classical probability flux for a Brownian particle is written as
\begin{eqnarray}
j_{cl} (x, t) &\equiv& u_{cl} (x,t) \, \rho_{cl} (x,t) = \left[ \frac{p_0}{m} ~e^{-2\ga t} +  \right.\nonumber \\
&+&  \left. \frac{D}{m^2} \frac{ x - x_t }{\big\langle(x(t)-x(0))^2\big\rangle_{cl}} f(t)^2 \right] \rho_{cl} (x, t) .
\end{eqnarray}

\subsection{Ballistic and diffusive motions}

In order to proceed further, one needs to invoke the two extreme time behaviors which are well established in surface diffusion; the ballistic motion which takes place when $t << \gamma^{-1}$, and the Brownian or diffusive motion when $t >> \gamma^{-1}$. 
Thus, at short times and by considering the linear term of the exponential, Eqs. (\ref{xt}) and (\ref{wt}) reduce to
\begin{equation}\label{quantum}
	x_t \approx v_0\, t,\hspace{1cm}
	\ti \si_{t}^2 \approx \si_0^2 \left( 1 + \frac{ \ti \hb^2 }{ 4\, m^2 \si_0^4 } t^2 \right)  ,
\end{equation}
and when approaching the classical limit to
\begin{equation}\label{classical}
	\si_{t}^2 \approx v_0^2 t^2 .
\end{equation}
Interestingly enough, the typical $t^2$-behavior valid at short times implies that the origin 
is substantially different in both dynamical regimes: quantum, Eq.(\ref{quantum}), and classical, Eq. (\ref{classical}). Furthermore, the motion of the Brownian particle is friction free as well as the generalized velocity $\ti u$  included in the flux defined by Eq (\ref{eq: cur}).
On the contrary, in the Brownian regime, one has that
\begin{equation}\label{long}
	x_t \approx \frac{v_0}{2\ga},\hspace{0.5cm}
	\ti \si_{t}^2 \approx \si_0^2 \left[ 1 + \frac{ \ti \hb^2 }{ 16\, m^2 \si_0^4 \ga^2 }  + \left(\frac{4\ga\, t}{8\, m^2\ga^3\si_0^2}\right)D\right]  ,
\end{equation}
that is, only the MSD $\langle(x(t)-x(0))^2\rangle$ survives. 
And again, approaching the classical limit, we have
\begin{equation}\label{longc}
\si_{t}^2 \approx  \Big\langle\big(x(t)-x(0)\big)^2\Big\rangle_{cl}  ,
\end{equation}
that is, no quantum behavior is observed.

\subsection{Quantum and classical response functions}

As mentioned in the Introduction, two response functions in this context are the ISF and the DSF which are Fourier transform one from the other. The ISF is the space Fourier transform of the
probability \cite{Torres2022,Torres2023} and, in the scaled theory, reads as follows
\begin{eqnarray}\label{ISFs}
	\ti I(K,t) &=& \int dx \, \, e^{- i x\, K} \, \ti \rho(x,t)  \nonumber \\
	 &=& \exp\left\{-\frac{1}{2} K \, (K\, \ti \si_t^2 - 2\, i\, x_t)\right\} ,
\end{eqnarray}
which is also Gaussian in the momentum transfer $K$. In the ballistic regime, one has that
\begin{equation}\label{Is_ballistic}
	\ti I^{ballistic}(K,t) \simeq \exp\left\{-\frac{1}{2} K^2 \si_0^2 + i\, K \,v_0\, t\right\}\exp\left\{-\frac{K^2 \ti \hb^2}{8\, m^2\si_0^2}t^2\right\},
\end{equation}
that is, a pure Gaussian function in time and momentum transfer is obtained. The time exponential factor depends on the particle mass and the initial width.  The scaled recoil energy of the adsorbate is $\ti E_r = K^2 \ti \hb^2/ 2\, m^2$. It is important to stress again at this point that the origin of the ballistic motion is different in the quantum regime, it comes from the commutator and gives rise to the adsorbate recoil energy, and the classical regime which comes from the MSD. When approaching the classical ISF, one has the same time and momentum behaviors but with the presence of the thermal velocity  
\begin{eqnarray}\label{Icl_ballistic}
I_{cl}^{ballistic}(K,t) &=& \exp\left\{-\frac{1}{2} K \, (K\, \Big\langle\big(x(t)-x(0)\big)^2 \Big\rangle_{cl} - 2\, i\, x_t)\right\} \nonumber \\
&\simeq& \exp\left\{ i\, K \,v_0\, t\right\}\exp\left\{-K^2\, v_0^2 \, t^2\right\}  .
\end{eqnarray}
%
%Furthermore, the generalized velocity $\ti u(x,t)$ has only one component which is given by 
%
%\begin{equation}
%\ti u	(x,t) \approx (x-x_t) \frac{D}{4 m^2 \ga^2 \ti \sigma_t^2} ,
%\end{equation}
%
%where $\ti \sigma_t$ is given by Eq.(\ref{long}). 
%In this regime, the general velocity is expressed as 
%two factors: the first one depends on position, 
%and the second one on time. 

An interesting aspect which worthwhile mentioning is the analysis of the full width at hall maximum (FWHM) in the time domain for the ballistic motion. The corresponding ISFs given by Eqs. (\ref{Is_ballistic}) and Eq.(\ref{Icl_ballistic}) can in general be expressed as a Gaussian function given by 
\begin{equation}\label{Igauss}
I(K,t) = \exp\left(- t^2/ 2 \sigma^2\right)  ,
\end{equation}
with
\begin{equation}\label{fwhm}
FWHM = 2 \, \hbar \sigma^{-1} \sqrt{2 \, \ln 2}  ,
\end{equation}
and where for each dynamical regime, one has 
\begin{equation} \label{sigma1}
\ti \sigma^{-1} = K \, \ti v_s \, , 
\end{equation}
for a quantum ideal gas and
\begin{equation}\label{sigma2}
\sigma_{cl}^{-1} = K \, v_0   ,
\end{equation}
for a classical ideal gas.
The distinction between quantum and classical ideal gas comes from the expression for $\sigma$ due to the different regime which is ruling the ballistic motion, quantum or classical motion, respectively.
These are important expressions since the $\sigma$-value is known from experiment. Eqs. (\ref{sigma1})  and (\ref{sigma2}) show us that the inverse of the $\sigma$-width is linear with the momentum transfer and the corresponding slopes are given by the velocities $\ti v_s$ and $v_0$, respectively.
In the scaled regime, this velocity depends on the adsorbate mass and the initial width of the wave packet 
whereas, in the classical case, it depends not only on the mass but also on the temperature through $v_0$.

\begin{figure}[t]
%    \setlength{\belowcaptionskip}{-15pt}
%	\centering
 %       \includegraphics[height=6.3cm]{Figure1.pdf}
        \includegraphics[height=6.3cm]{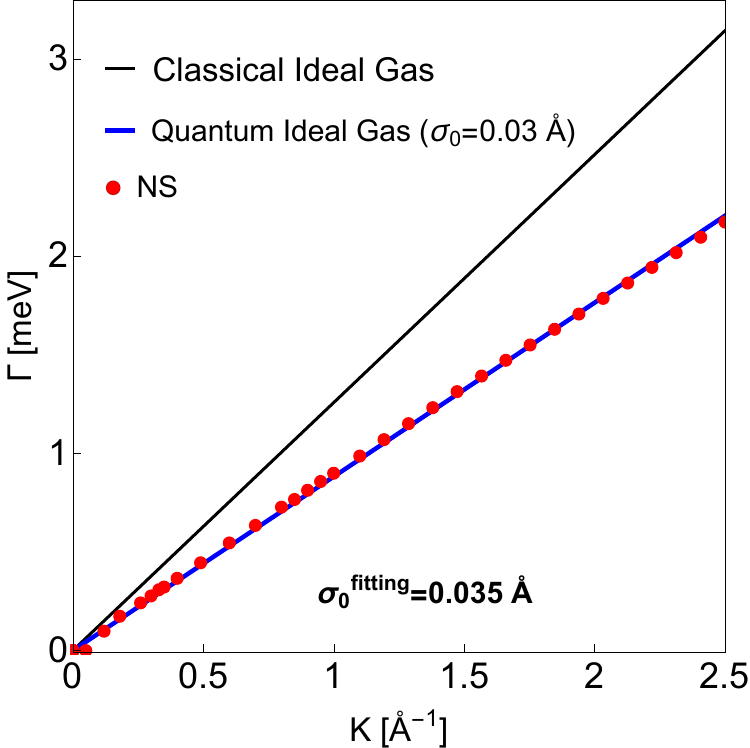}
	\caption{FWHM as a function of momentum transfer $K$ for Xe adsorbates on Pt(111) in the ballistic regime. Black solid line corresponds to a classical ideal gas from Eqs. (\ref{Igauss}),  (\ref{fwhm}) and (\ref{sigma2}). Numerical simulations (NS) are displayed by red points which are issued from the SWF method for a potential barrier of $9.6\,\text{meV}$, friction coefficient $\gamma = 0.05\,\text{ps}^{-1}$, incident energy of $10.15\,\text{meV}$, $T=105 \, K$ and incident width of the Gaussian wave packet $\sigma_0= 0.03 \, \AA$. Blue line corresponds to a fitting from Eq. (\ref{sigma1}) (quantum ideal gas) with $\sigma_0= 0.035 \, \AA$.}
	\label{Ballistic-results}
\end{figure}

As an illustration, let us consider the paradigmatic example of Xe on a Pt(111) surface. 
Fig. \ref{Ballistic-results} shows the dependence of  $\Gamma$ versus the momentum transfer $K$. 
Black solid line corresponds to a classical ideal gas issued from Eqs. (\ref{Igauss}), (\ref{fwhm}) and 
(\ref{sigma2}). Red points are numerical simulations (NS) obtained from the SWF method for a potential 
barrier of $9.6\,\text{meV}$, friction coefficient $\gamma = 0.05\,\text{ps}^{-1}$, $T=105 \, K$, incident energy 
of $10.15\,\text{meV}$ and incident width of the Gaussian wave packet 
$\sigma_0= 0.03 \, \AA$. \cite{Torres2023} Finally, the blue line corresponds to a fitting from 
Eq. (\ref{sigma1}) (quantum ideal gas) with $\sigma_0= 0.035 \, \AA$. This confirms us that the
numerical simulations are also displaying a straight line which agrees with the theoretical prediction 
given by Eq. (\ref{sigma1}).
This deviation observed between the two slopes have been usually attributed to an effective mass when considering the surface diffusion.  This is due exclusively to the role played by the width of the wave packet characterizing the adsorbate in this quantum dynamics.
Thus, in the ballistic regime, where the motion is free, one can not speak of effective mass for the adsorbates; only when the interaction with the surface is important (beyond the ballistic regime).

On the other hand, in the Brownian regime, the scaled ISF is expressed as 
\begin{eqnarray}\label{Is_brownian}
	\ti I^{Brownian}(K,t) &=& \exp \Bigg[-\frac{1}{2}K^2\left(
	\si_0^2 + \frac{\ti \hb^2}{16\, m^2\si_0^2\ga^2} - \frac{3 D}{8 \, m^2\ga^3} \right) \nonumber \\
	&+& i\, K\frac{v_0}{2\ga}\Bigg]
	\times\exp\left\{-\frac{K^2\,D}{4\, m^2\ga^2}t\right\}.
\end{eqnarray}
As expected, the scaled ISF gradually passes from a Gaussian to an exponential function in time when going from the short time to the long time behaviors. For the classical limit, the ISF reads as
\begin{equation}\label{Icl_brownian}
I_{cl}^{Brownian}(K,t) = \exp\left\{\frac{3 K^2 D}{16 \, m^2\ga^3}  
+ i\, K\frac{v_0}{2\ga} \right\}
\exp\left\{-\frac{K^2\,D}{4\, m^2\ga^2}t\right\} ,
\end{equation}
which has the same exponential behavior in time than for the scaled ISF. The decay rate or dephasing rate, $\alpha (K)$ depends on the momentum transfer (quadratic), friction, mass and surface temperature through the diffusion constant. Finally, the DSF is then expressed as
\begin{equation}\label{lorentzian-free}
S(K,\omega) = \frac{\alpha(K)}{\omega^2 + \alpha(K)^2} , \hspace{1cm} \alpha(K)= \frac{K^2\,D}{4\, m^2\ga^2} .
\end{equation}
The diffusion coefficient can also be extracted from the long time limit of the quantum or classical position variance
\begin{equation}\label{Eq: D-Einstein}
	D= \lim_{t\rightarrow \infty} \frac{1}{4\,t}\Big\langle\,\big(x(t)- x(0)\big)^2\Big\rangle_{cl} .
\end{equation}

\section{Master equation and H-theorem for surface diffusion}

As mentioned above, the ISF, $I(\textbf{K},t)$, is expressed in terms of the probability density  
$\rho({\bf R},t)$  as follows \cite{torres2021surface, Torres2022,Torres2023} 
\begin{equation}\label{ISF11}            
I({\bf K},t)  =  \int d {\bf R} \, e^{i {\bf K} \cdot {\bf R}} \, \rho({\bf R},t) .
\end{equation}
%
%
%In the ballistic time regime, the motion of the particle is free and thus a Gaussian ansatz for the probability in this regime 
%is clearly justified. The same discussion found in 
%previous Section can be extended here. The origin of the $t^2$-behavior of the width and ISF is different if the quantum or classical regime is considered.
Recently, we have proposed that the ISF can also be seen as a characteristic function of probability theory.\cite{Torres2025a}

As we know, the diffusive behavior is reached at very long times, when time is much greater than the inverse of the friction coefficient. At long times, the ISF can be fitted to an exponential function of time according to
\begin{equation}\label{ISF2}
I({\bf K}, t) = B \, e^{-   \alpha({\bf K}) \, t} + C,
\end{equation}
where $B$ and $C$ are constants and $\alpha({\bf K})$ is the so-called dephasing rate. 
The $\alpha$-rate is also obtained by solving the Markovian master equation and valuable physical information of the system under study can be extracted from this rate by assuming a jump diffusion model: adsorbate-substrate and adsorbate-adsorbate interactions, friction coefficients, surface corrugation as well as classical/quantum hopping rates.

In general, the surface diffusion is an activated process and, depending on the nature of the adsorbate and surface temperature, could also be assisted by tunneling. One of the most important physical quantities is the total jumping rate which follows a typical Arrhenius plot at high temperatures (when plotting the log of this magnitude versus the inverse of the temperature, a straight line is observed). When reaching the so-called crossover temperature, where quantum diffusion by tunneling is predominant, there is a change of slope. In this dynamics, the role played not only by the friction and 
activation energy is important but also the surface structure. 
This is emphasized in the theory when the dephasing rate is written as a function of $\textbf K$. All of these physical parameters are able to be known if a theory exists in order to interpret and consistently support experimental data. When the diffusive time regime is established, what remains important is the probability of jumping due to thermal activation or tunneling. The corresponding probabilities and thus the jumping rates have to be fitted or calculated in order to cover in a continuous way the full range going from high to small surface temperatures.

First of all, let us briefly remind the jump diffusion CE model.\cite{Chudley}
If, for simplicity, a simple Bravais lattice is assumed as well as instantaneous jumps between different sites, the Pauli master equation in space and time (which is also a quantum and classical kinetic equation) can be written as
\begin{equation}\label{Pauli}
\frac{\partial \rho (\textbf{R},t)}{\partial t} = \sum_{\bf L} \Gamma_{\bf L} \big[\rho(\textbf{R}+{\bf L},t) - \rho(\textbf{R},t)\big]  ,
\end{equation}
where summation runs over all two-dimensional vectors ${\bf L}$ 
and is written in terms of the partial jumping rates $\Gamma_{{\bf L}}$; $\Gamma_{{\bf L}}^{-1}$ represents the average time between successive jumps. It is also assumed
that the time for a simple jump is very short compared with the time between successive jumps. The total jump rate is therefore $\Gamma = \sum_{{\bf L}} \Gamma_{{\bf L}}$, with  $\Gamma_{{\bf L}} = \Gamma_{-{\bf L}}$. 

One way of solving Eq. (\ref{Pauli}) is to take advantage of the linearity property of the Fourier transform. Thus, according to Eq.(\ref{ISF1}), one has the following differential equation in terms of the ISF
\begin{equation}
\frac{\partial I ({\bf K},t)}{\partial t} = - 2 I({\bf K},t) \sum_{\bf L>0} \Gamma_{\bf L} \big( 1 - \cos({\bf L}\cdot  {\bf K}) \big),
\end{equation}
which solution is simply given by 
\begin{equation}\label{eq:Ialfa}
I({\bf K},t) = I({\bf K},0) e^{- 2 | t | \sum_{\bf L>0} \Gamma_{\bf L} \big( 1 - \cos \left({\bf L} \cdot {\bf K}\right) \big)} ,
\end{equation}
that is, the effect of including a simple Bravais lattice is to add an exponential factor ruling the jumps in the activated diffusion process. The dephasing rate is thus written as
\begin{equation}\label{eq:alpha}
\alpha({\bf K}) = 2 \, \Gamma \, \sum_{\bf L > 0} P_{\bf L} \big( 1 - \cos\left({\bf L} \cdot {\bf K}\right) \big) ,
\end{equation}
where the jump probabilities are given by $P_{{\bf L}} = \Gamma_{{\bf L}} / \Gamma$. From the experimental dephasing rate obtained from Eq. (\ref{ISF2}), the fitting parameters in this model are the partial jumping rates, $\Gamma_{{\bf L}} $. Thus, for example, if only one-dimensional surface direction is considered, the sum over $l$ can 
be replaced by an integral for values of $l \geq 1$ and by assuming
an exponential function for the jump probability $P_l \simeq e^{-b l}$, then the dephasing rate is periodic and expressed as \cite{Guantes2002}
\begin{equation}\label{alpha-gamma}
\alpha(K) =  \Gamma \, \frac{2 \, \big(1+e^b\big) \big(1 + \cos(a K)\big)}{\big(e^b - 1\big) \big(\cosh \,b + \cos (a K)\big)}   ,
\end{equation}
$a$ being the unit cell length. Eq. (\ref{alpha-gamma}) is 
interesting because it tells us that the dephasing rate and the total jump rate are proportional except for a factor which depends on $K$ (the momentum transfer along the surface), $a$ and $b$. Obviously, $\Gamma$ is an implicit function of the surface temperature and friction. Now, if $P_n = \rho_{nn}$ is the probability to stay at the $n$th-well at time t then
\begin{eqnarray}\label{Ik1}
	P_n(t) &=& I(K,0) \sum_{K}  e^{- |t| \alpha(K)} \, e^{-i K n}  .
\end{eqnarray}

Now, if only transitions between first neighbors are going to be allowed (the diffusion dynamics is assumed to be one-dimensional on a periodic substrate, along a symmetry direction), the probability to stay at the $n$th-well at time $t$, $P_n (t)= \rho_{nn}(t)$ can be written as \cite{Ruth2013,Elena2025}
%
%\begin{equation}\label{Pdot}
%	\dot P_n (t) = \Gamma^+_{n-1} P_{n-1}(t) +  \Gamma^-_{n+1} P_{n+1} (t) - (   %\Gamma^+_{n} + \Gamma^-_{n} )  P_{n}(t)  ,
%\end{equation}
% 
%with  being the probability to stay at the $n$th-well, $\Gamma^{\pm}_{n \pm 1}$ the tunneling/hopping transition rates from the $(n \mp 1)$-th well to the $n$-th well and $\pm$ denotes if diffusion goes to the right or to the left, respectively. If the initial condition is such that $P_n(0) = \delta_{n0}$ and $\Gamma = \Gamma^+ + \Gamma^-$ describes the total rate (with $\Gamma^+_{n} = \Gamma^-_{n}$,  $\Gamma^+_{n} = \Gamma^+$, and $\Gamma^-_{n} = \Gamma^-$), then the solution is given by \cite{Elena2025}
%
\begin{equation}\label{Psol}
	P_n (t) = I_n(\Gamma t) \, e^{- \Gamma t},
\end{equation}
where $I_n(x)$ is the modified Bessel function of integer order $n$.  Thus, $P_n(t)$ gives us then the probability to stay in the $n$th-well of the binding site at time $t$. The ISF is then given by
\begin{eqnarray}\label{Ik2}
	I (K,t) &=& \sum_{n} P_n(t) \, e^{i K n}   \nonumber \\
                & = &e^{- \Gamma  (1-cos \, K) t} \nonumber \\
                &= & e^{- \Gamma t} \sum_{n=- \infty}^{+ \infty} I_n(\Gamma t) e^{i n K}   .
\end{eqnarray}
% 
%where $K_{||} = K a \cos(\beta)$, $\beta$ being the angle formed by the direction of observation and diffusion symmetry direction
%(in our case, $\beta = \pi / 6$ for the two symmetry directions allowed by $\text{\bf K}$). 
%$K_{||}$ is then the momentum transfer along the projection on the one dimensional lattice/direction with lattice points labeled by $n$. 
%
%and 
%
%\begin{equation}\label{al1}
%\Gamma = \frac{\alpha (K)}  {1-\cos(K a \cos(\pi/6))} .
%\end{equation}

As pointed out before, the main drawback of the CE model is that, in practice, the rates $\Gamma_{\textbf {L}}$ are fitting parameters and not known {\it a priori}. It is here that Kramers' theory
\cite{Eli2023} provides analytical expressions for jump classical rates and probabilities,
based on knowledge of a single parameter, namely the reduced energy loss as the system 
moves from one barrier to the next. From this theoretical treatment, in addition to frictions, 
barrier heights and well and barrier frequencies can be easily extracted since they are acting as fitting parameters. The turnover 
theory as applied to surface diffusion is also based on a master equation (similar to the CE model) for the populations 
in the wells, governed by an energy exchange kernel. This theory comes initially from 
chemical reactions and applied to the reaction coordinate. It is basically a 
one-dimensional theory. This formalism has been successfully applied to 
the diffusion of Na atoms on a metallic surface. \cite{Vega2002,Guantes2003} The total 
jump classical rate, $\Gamma$, follows an Arrhenius plot when plotting versus the inverse of the surface 
temperature.

It is very important to emphasize that this theory is suitable for one-dimensional coordinate which is chosen to be a symmetric direction of the surface where diffusion takes place. The resulting theory is analytic if one approximates the reflection and transmission coefficients using the parabolic barrier expressions and by assuming separable motion along the symmetry surface direction  between the so-called unstable normal mode and the stable modes. This is an important assumption which could delimit some times the application of this theory.\cite{Eli2023,Elena2025}

When a master equation is assumed to be applicable to a given problem, it is always possible to define an H-function as \cite{Liboff}
\begin{equation}\label{H1}
H = \sum_n P_n \ln P_n
\end{equation}
and where the entropy is written in terms of the H-function  as 
\begin{equation}\label{S}
S = - k_B H ,
\end{equation}
$k_B$ being the Boltzmann constant. The H-theorem establishes that 
\begin{equation}\label{H2}
\frac{d H}{d t} \leq 0  ,
\end{equation}
and, therefore, the system entropy always increases with time.

The diffusion of H and D on a Pt(111) surface is paradigmatic here since, thanks to the surface temperature, this surface diffusion moves from  thermal activation to tunneling, thus monitoring the quantum-to-classical transition within the same system.
The corresponding measurements in the range of surface temperatures used went from 250 K up to 80 K, covering thermal activation and tunneling regimes.\cite{jardine2010determination} The crossover temperature was estimated to be 66 K for H and 63 K for D. Furthermore, from an experimental CE analysis, the diffusion motion was established to correspond to nearest neighbor hopping for a coverage of $0.1$ ML, along the $[11\bar2]$ direction. and with a lattice length of $a=2.77 \, \text{\AA}$. Deviations from nearest neighbor random jumps for H and D between fcc hollow sites of the Pt(111) surface were reported to be minimal. Incoherent surface tunneling diffusion has been studied by solving the corresponding It\^o stochastic differential equation.\cite{Elena2025}
In this calculation, a parabolic barrier has been assumed with a barrier value of $V^\ddag = 72 \, meV$ for H and D, well and barrier frequencies given by $\hbar \omega_0 = \hbar \omega^\ddag= 30 \, \, meV$ for H; the corresponding frequency for D is given by  $\hbar \omega_0 / \sqrt{2}$ due to the relation between masses. And, finally, if Ohmic friction is also assumed the corresponding values are 140 $ps^{-1}$ and 70 $ps^{-1}$ for H and D, respectively. These high values could explain why diffusion through first neighbor sites are prominent. For H diffusion, with a surface temperature of $214$ K, hops by thermal activation dominate, $100-90$\%; on the contrary, for $80$ K, close to the crossover temperature which is around $60$ K, only hops by tunneling predominates, around $90$\%; and, finally, for $121$ K, hops for thermal activation and tunneling are around $30$\% and $70$\%, respectively.\cite{Elena2025}
The hopping rates for D follow the same tendency more or less but are consistently smaller than those for H across the entire temperature range, indicating a clear mass effect.

The Pauli master equation is given by Eq. (\ref{Pauli}) and the corresponding solution is expressed in Eq. (\ref{Psol}). The $\Gamma$-value which represents the total jump/hopping rate at a given surface temperature is chosen to be that provided by the experiment work.\cite{jardine2010determination} 
Thus, according to Eq. (\ref{H1}), in our example, the $H$-function is written as
\begin{equation}\label{eq:H-function}
H (t) = e^{- \Gamma t} \sum_n I_n(\Gamma t) \, \big[ \ln I_n(\Gamma t)  - \Gamma t \big]  .
\end{equation}
By defining $z= \Gamma t$, then the time derivative of H 
in terms of the new variable $z$ is now expressed as
\begin{equation}
\frac{\partial H (z)}{\partial z} =  \Gamma e^{- z} \sum_n \left[\frac{I_{n-1}(z) + I_{n+1}(z)}{2}-I_n(z) \right] \, \left( \ln I_n(z)  - z \right) \leq 0
\end{equation}
since the derivative of the argument of the modified Bessel function can be written in terms of the semi-sum of the same functions but of order $n-1$ and $n+1$.\cite{Table}

\begin{figure}[t]
    \centering
    \includegraphics[width=0.70\linewidth]{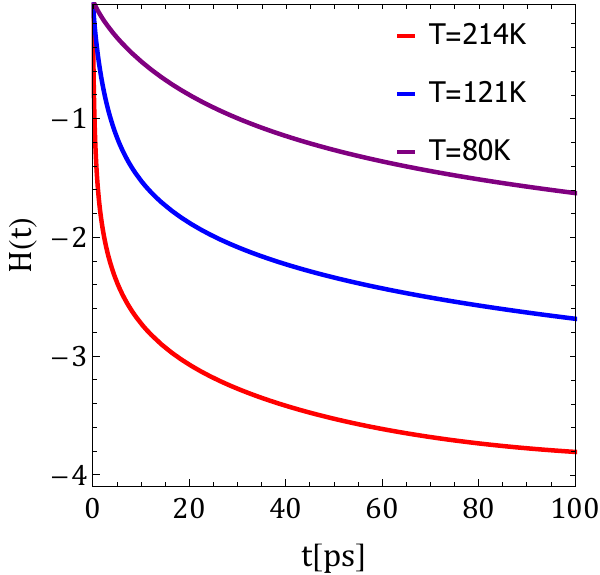}
    \caption{The H-function from Eq. (\ref{eq:H-function}) is plotted versus time for H adsorbates on a Pt(111) surface at three different temperatures: $214\, \text{K}$ (red curve), $121\, \text{K}$ (blue curve) and $80\, \text{K}$ (purple curve). These calculations are carried out using the corresponding experimental hopping/tunneling rates obtained for a momentum transfer of $K= 0.86\, \textup{~\AA}^{-1}$ along the $[11\bar2]$ direction and a coverage of 0.1 ML.}
    \label{F1}
\end{figure}

The $H$-function is plotted versus time in Fig \ref{F1} in order to show how this time derivative behaves for three different regimes, pure classical, pure quantum and in-between. The only parameter in this kind of plots is the total jump rate, $\Gamma$, which is an explicit function of the surface temperature according to an Arrhenius-like plot. This Arrhenius-like behavior can be extracted from the experimental data (where a factor $2/3$ has been considered due to the two symmetry directions allowed by the direction of observation $[11\bar2]$), from the solution of the It\^o stochastic differential equation or from Kramers' theory (when applicable).\cite{Elena2025} It is clear that tunneling delays irreversibility versus the thermal activation diffusion occurring at the highest surface temperature. In fact, we have recently reported that for $214\, \text{K}$ (red curve) diffusion is fully mediated by thermal activation, $121\, \text{K}$ (blue curve) is around fifty-fifty and $80\, \text{K}$ (purple curve) is mainly tunneling diffusion.

\section{Conclusions}
Along this work, we have put in evidence how the quantum-to-classical transition can proceed for an open dynamics: the surface diffusion. Two different ways to study this transition have been used, through a general scaled theory of Brownian motion and varying the surface temperature in order to cover the thermal activation and incoherent tunneling diffusion. In the first part, the scaled CL master equation is used showing that the origin of the ballistic motion (short times) is quite different when approaching in a continuous and smooth way to the classical regime in spite of the fact that the $t^2$-dependence of the ISF is kept along this transition. A discussion on the so-called effective mass in this context is also presented emphasizing the fact that this concept should not be used since in the ballistic regime, diffusion is frictionless. Even more, thanks to this type of analysis one can speak of classical and quantum ideal gas. In the diffusive regime (long times), no trace of quantum features along this transition remains. In order to process the experimental results, a master equation is usually considered within the CE model or Krames`theory.  When dealing with Pauli-like master equation is quite straightforward to carry out an analysis of the well-known H-function. A detailed study of this function has been presented for three different surface temperatures which clearly illustrate the quantum-to-classical transition for the H and D diffusion on a Pt(111) surface. Irreversibility is clearly delayed by tunneling surface diffusion and, therefore, one has a way to discriminate classical from pure quantum behavior.

%
%\appendix{}

%\section{Gauss....}\label{appendix A}
%

%
\section*{Author Contributions}
All authors have contributed equally to this work.

\section*{Conflicts of interest}
There are no conflicts to declare.

\section*{Acknowledgments}
E.E.T.-M. and S.M.-A. would like to thank W. Allison, D. J. Ward and M.Ord for very stimulating discussions. They also acknowledge support of a grant from the Ministry of Science, Innovation and Universities with Ref. PID2023-149406NB-I00.

%%%END OF MAIN TEXT%%%


%\balance


%%%REFERENCES%%%
\bibliography{rsc.bib} %You need to replace "rsc" on this line with the name of your .bib file

\providecommand*{\mcitethebibliography}{\thebibliography}
\csname @ifundefined\endcsname{endmcitethebibliography}
{\let\endmcitethebibliography\endthebibliography}{}
\begin{mcitethebibliography}{52}
\providecommand*{\natexlab}[1]{#1}
\providecommand*{\mciteSetBstSublistMode}[1]{}
\providecommand*{\mciteSetBstMaxWidthForm}[2]{}
\providecommand*{\mciteBstWouldAddEndPuncttrue}
  {\def\EndOfBibitem{\unskip.}}
\providecommand*{\mciteBstWouldAddEndPunctfalse}
  {\let\EndOfBibitem\relax}
\providecommand*{\mciteSetBstMidEndSepPunct}[3]{}
\providecommand*{\mciteSetBstSublistLabelBeginEnd}[3]{}
\providecommand*{\EndOfBibitem}{}
\mciteSetBstSublistMode{f}
\mciteSetBstMaxWidthForm{subitem}
{(\emph{\alph{mcitesubitemcount}})}
\mciteSetBstSublistLabelBeginEnd{\mcitemaxwidthsubitemform\space}
{\relax}{\relax}

\bibitem[Hove(1954)]{vanHove}
L.~V. Hove, \emph{Physical Review}, 1954, \textbf{95}, 249\relax
\mciteBstWouldAddEndPuncttrue
\mciteSetBstMidEndSepPunct{\mcitedefaultmidpunct}
{\mcitedefaultendpunct}{\mcitedefaultseppunct}\relax
\EndOfBibitem
\bibitem[Vineyard(1958)]{Vineyard}
G.~Vineyard, \emph{Phys. Rev.}, 1958, \textbf{110}, 999\relax
\mciteBstWouldAddEndPuncttrue
\mciteSetBstMidEndSepPunct{\mcitedefaultmidpunct}
{\mcitedefaultendpunct}{\mcitedefaultseppunct}\relax
\EndOfBibitem
\bibitem[Lovesey(1984)]{Lovesey}
S.~Lovesey, \emph{``{T}heory of neutron scattering from condensed matter. Vol.
  2. {P}olarization effect and magnetic scattering''}, Clarendon Press,
  1984\relax
\mciteBstWouldAddEndPuncttrue
\mciteSetBstMidEndSepPunct{\mcitedefaultmidpunct}
{\mcitedefaultendpunct}{\mcitedefaultseppunct}\relax
\EndOfBibitem
\bibitem[McQuarrie(1976)]{mcquarrie1976statistical}
D.~McQuarrie, \emph{Statistical Mechanics, Chap. 21}, Harper and Row, New York,
  1976\relax
\mciteBstWouldAddEndPuncttrue
\mciteSetBstMidEndSepPunct{\mcitedefaultmidpunct}
{\mcitedefaultendpunct}{\mcitedefaultseppunct}\relax
\EndOfBibitem
\bibitem[Hofmann and Toennies(1996)]{Hofman1996}
F.~Hofmann and J.~P. Toennies, \emph{Chemical Reviews}, 1996, \textbf{96},
  1307--1326\relax
\mciteBstWouldAddEndPuncttrue
\mciteSetBstMidEndSepPunct{\mcitedefaultmidpunct}
{\mcitedefaultendpunct}{\mcitedefaultseppunct}\relax
\EndOfBibitem
\bibitem[Jardine \emph{et~al.}(2009)Jardine, Alexandrowicz, Hedgeland, Allison,
  and Ellis]{Jardine2009a}
A.~P. Jardine, G.~Alexandrowicz, H.~Hedgeland, W.~Allison and J.~Ellis,
  \emph{Phys. Chem. Chem. Phys.}, 2009, \textbf{11}, 3355--3374\relax
\mciteBstWouldAddEndPuncttrue
\mciteSetBstMidEndSepPunct{\mcitedefaultmidpunct}
{\mcitedefaultendpunct}{\mcitedefaultseppunct}\relax
\EndOfBibitem
\bibitem[Jardine \emph{et~al.}(2009)Jardine, Hedgeland, Alexandrowicz, Allison,
  and Ellis]{Jardine2009b}
A.~Jardine, H.~Hedgeland, G.~Alexandrowicz, W.~Allison and J.~Ellis,
  \emph{Progress in Surface Science}, 2009, \textbf{84}, 323--379\relax
\mciteBstWouldAddEndPuncttrue
\mciteSetBstMidEndSepPunct{\mcitedefaultmidpunct}
{\mcitedefaultendpunct}{\mcitedefaultseppunct}\relax
\EndOfBibitem
\bibitem[McQuarrie(1976)]{mcquarrie76a}
D.~McQuarrie, \emph{``{S}tatistical {M}echanics''}, Harper Collins, New York,
  1976\relax
\mciteBstWouldAddEndPuncttrue
\mciteSetBstMidEndSepPunct{\mcitedefaultmidpunct}
{\mcitedefaultendpunct}{\mcitedefaultseppunct}\relax
\EndOfBibitem
\bibitem[Zwanzig(2001)]{Zwanzig2001}
R.~Zwanzig, \emph{Nonequilibrium statistical mechanics}, Oxford university
  press, 2001\relax
\mciteBstWouldAddEndPuncttrue
\mciteSetBstMidEndSepPunct{\mcitedefaultmidpunct}
{\mcitedefaultendpunct}{\mcitedefaultseppunct}\relax
\EndOfBibitem
\bibitem[Schiller(1962)]{Schiller1}
R.~Schiller, \emph{Physical Review}, 1962, \textbf{125}, 1100\relax
\mciteBstWouldAddEndPuncttrue
\mciteSetBstMidEndSepPunct{\mcitedefaultmidpunct}
{\mcitedefaultendpunct}{\mcitedefaultseppunct}\relax
\EndOfBibitem
\bibitem[Schiller(1962)]{Schiller2}
R.~Schiller, \emph{Physical Review}, 1962, \textbf{125}, 1109\relax
\mciteBstWouldAddEndPuncttrue
\mciteSetBstMidEndSepPunct{\mcitedefaultmidpunct}
{\mcitedefaultendpunct}{\mcitedefaultseppunct}\relax
\EndOfBibitem
\bibitem[Schiller(1962)]{Schiller3}
R.~Schiller, \emph{Physical Review}, 1962, \textbf{125}, 1116\relax
\mciteBstWouldAddEndPuncttrue
\mciteSetBstMidEndSepPunct{\mcitedefaultmidpunct}
{\mcitedefaultendpunct}{\mcitedefaultseppunct}\relax
\EndOfBibitem
\bibitem[Rosen(1964)]{Rosen1}
N.~Rosen, \emph{American Journal of Physics}, 1964, \textbf{32}, 597--600\relax
\mciteBstWouldAddEndPuncttrue
\mciteSetBstMidEndSepPunct{\mcitedefaultmidpunct}
{\mcitedefaultendpunct}{\mcitedefaultseppunct}\relax
\EndOfBibitem
\bibitem[Rosen(1986)]{Rosen2}
N.~Rosen, \emph{Foundations of physics}, 1986, \textbf{16}, 687--700\relax
\mciteBstWouldAddEndPuncttrue
\mciteSetBstMidEndSepPunct{\mcitedefaultmidpunct}
{\mcitedefaultendpunct}{\mcitedefaultseppunct}\relax
\EndOfBibitem
\bibitem[Holland(1995)]{Holland}
P.~R. Holland, \emph{The quantum theory of motion: an account of the de
  Broglie-Bohm causal interpretation of quantum mechanics}, Cambridge
  university press, 1995\relax
\mciteBstWouldAddEndPuncttrue
\mciteSetBstMidEndSepPunct{\mcitedefaultmidpunct}
{\mcitedefaultendpunct}{\mcitedefaultseppunct}\relax
\EndOfBibitem
\bibitem[Sanz and Miret-Art{\'e}s(2012)]{Salva-b1}
{\'A}.~S. Sanz and S.~Miret-Art{\'e}s, \emph{A Trajectory Description of
  Quantum Processes. I. Fundamentals: A Bohmian Perspective}, Springer, 2012,
  vol. 850\relax
\mciteBstWouldAddEndPuncttrue
\mciteSetBstMidEndSepPunct{\mcitedefaultmidpunct}
{\mcitedefaultendpunct}{\mcitedefaultseppunct}\relax
\EndOfBibitem
\bibitem[Sanz and Miret-Art{\'e}s(2013)]{Salva-b2}
{\'A}.~S. Sanz and S.~Miret-Art{\'e}s, \emph{A Trajectory Description of
  Quantum Processes. II. Applications: A Bohmian Perspective}, Springer, 2013,
  vol. 831\relax
\mciteBstWouldAddEndPuncttrue
\mciteSetBstMidEndSepPunct{\mcitedefaultmidpunct}
{\mcitedefaultendpunct}{\mcitedefaultseppunct}\relax
\EndOfBibitem
\bibitem[Richardson \emph{et~al.}(2014)Richardson, Schlagheck, Martin,
  Vandewalle, and Bastin]{Richardson2014}
C.~D. Richardson, P.~Schlagheck, J.~Martin, N.~Vandewalle and T.~Bastin,
  \emph{Physical Review A}, 2014, \textbf{89}, 032118\relax
\mciteBstWouldAddEndPuncttrue
\mciteSetBstMidEndSepPunct{\mcitedefaultmidpunct}
{\mcitedefaultendpunct}{\mcitedefaultseppunct}\relax
\EndOfBibitem
\bibitem[Chou(2016)]{Chou1}
C.-C. Chou, \emph{Annals of Physics}, 2016, \textbf{371}, 437--459\relax
\mciteBstWouldAddEndPuncttrue
\mciteSetBstMidEndSepPunct{\mcitedefaultmidpunct}
{\mcitedefaultendpunct}{\mcitedefaultseppunct}\relax
\EndOfBibitem
\bibitem[Chou(2016)]{Chou2}
C.-C. Chou, \emph{International Journal of Quantum Chemistry}, 2016,
  \textbf{116}, 1752--1762\relax
\mciteBstWouldAddEndPuncttrue
\mciteSetBstMidEndSepPunct{\mcitedefaultmidpunct}
{\mcitedefaultendpunct}{\mcitedefaultseppunct}\relax
\EndOfBibitem
\bibitem[Mousavi and Miret-Art{\'e}s(2024)]{Mousavi2024}
S.~Mousavi and S.~Miret-Art{\'e}s, \emph{The European Physical Journal Plus},
  2024, \textbf{139}, 926\relax
\mciteBstWouldAddEndPuncttrue
\mciteSetBstMidEndSepPunct{\mcitedefaultmidpunct}
{\mcitedefaultendpunct}{\mcitedefaultseppunct}\relax
\EndOfBibitem
\bibitem[Nassar and Miret-Art{\'e}s(2017)]{Nassar}
A.~B. Nassar and S.~Miret-Art{\'e}s, \emph{Bohmian mechanics, open quantum
  systems and continuous measurements}, Springer, 2017\relax
\mciteBstWouldAddEndPuncttrue
\mciteSetBstMidEndSepPunct{\mcitedefaultmidpunct}
{\mcitedefaultendpunct}{\mcitedefaultseppunct}\relax
\EndOfBibitem
\bibitem[Mousavi and Miret-Art{\'e}s(2022)]{Mousavi2022a}
S.~Mousavi and S.~Miret-Art{\'e}s, \emph{Foundations of Physics}, 2022,
  \textbf{52}, 78\relax
\mciteBstWouldAddEndPuncttrue
\mciteSetBstMidEndSepPunct{\mcitedefaultmidpunct}
{\mcitedefaultendpunct}{\mcitedefaultseppunct}\relax
\EndOfBibitem
\bibitem[Liboff(2003)]{Liboff}
R.~L. Liboff, \emph{Kinetic theory: classical, quantum, and relativistic
  descriptions}, Springer Science \& Business Media, 2003\relax
\mciteBstWouldAddEndPuncttrue
\mciteSetBstMidEndSepPunct{\mcitedefaultmidpunct}
{\mcitedefaultendpunct}{\mcitedefaultseppunct}\relax
\EndOfBibitem
\bibitem[Mousavi and Miret-Art{\'e}s(2023)]{Mousavi2023}
S.~V. Mousavi and S.~Miret-Art{\'e}s, \emph{Symmetry}, 2023, \textbf{15},
  1184\relax
\mciteBstWouldAddEndPuncttrue
\mciteSetBstMidEndSepPunct{\mcitedefaultmidpunct}
{\mcitedefaultendpunct}{\mcitedefaultseppunct}\relax
\EndOfBibitem
\bibitem[Torres-Miyares \emph{et~al.}(2022)Torres-Miyares, Rojas-Lorenzo,
  Rubayo-Soneira, and Miret-Art{\'e}s]{Torres2022}
E.~Torres-Miyares, G.~Rojas-Lorenzo, J.~Rubayo-Soneira and S.~Miret-Art{\'e}s,
  \emph{Physical Chemistry Chemical Physics}, 2022, \textbf{24},
  15871--15890\relax
\mciteBstWouldAddEndPuncttrue
\mciteSetBstMidEndSepPunct{\mcitedefaultmidpunct}
{\mcitedefaultendpunct}{\mcitedefaultseppunct}\relax
\EndOfBibitem
\bibitem[Torres-Miyares \emph{et~al.}(2023)Torres-Miyares, Ward, Rojas-Lorenzo,
  Rubayo-Soneira, Allison, and Miret-Art{\'e}s]{Torres2023}
E.~Torres-Miyares, D.~Ward, G.~Rojas-Lorenzo, J.~Rubayo-Soneira, W.~Allison and
  S.~Miret-Art{\'e}s, \emph{Physical Chemistry Chemical Physics}, 2023,
  \textbf{25}, 6225--6231\relax
\mciteBstWouldAddEndPuncttrue
\mciteSetBstMidEndSepPunct{\mcitedefaultmidpunct}
{\mcitedefaultendpunct}{\mcitedefaultseppunct}\relax
\EndOfBibitem
\bibitem[Caldeira and A.J.(1983)]{Caldeira1983}
A.~Caldeira and L.~A.J., \emph{Physica A}, 1983, \textbf{121}, 374\relax
\mciteBstWouldAddEndPuncttrue
\mciteSetBstMidEndSepPunct{\mcitedefaultmidpunct}
{\mcitedefaultendpunct}{\mcitedefaultseppunct}\relax
\EndOfBibitem
\bibitem[Lindblad(1976)]{Lindblad}
G.~Lindblad, \emph{Communications in Mathematical Physics}, 1976, \textbf{48},
  119\relax
\mciteBstWouldAddEndPuncttrue
\mciteSetBstMidEndSepPunct{\mcitedefaultmidpunct}
{\mcitedefaultendpunct}{\mcitedefaultseppunct}\relax
\EndOfBibitem
\bibitem[Khani \emph{et~al.}(2021)Khani, Mousavi, and
  Miret-Art{\'e}s]{Mousavi2021}
Z.~Khani, S.~V. Mousavi and S.~Miret-Art{\'e}s, \emph{Entropy}, 2021,
  \textbf{23}, 1469\relax
\mciteBstWouldAddEndPuncttrue
\mciteSetBstMidEndSepPunct{\mcitedefaultmidpunct}
{\mcitedefaultendpunct}{\mcitedefaultseppunct}\relax
\EndOfBibitem
\bibitem[Pollak and Miret-Art{\'e}s(2023)]{Eli2023}
E.~Pollak and S.~Miret-Art{\'e}s, \emph{ChemPhysChem}, 2023, \textbf{24},
  e202300272\relax
\mciteBstWouldAddEndPuncttrue
\mciteSetBstMidEndSepPunct{\mcitedefaultmidpunct}
{\mcitedefaultendpunct}{\mcitedefaultseppunct}\relax
\EndOfBibitem
\bibitem[Jardine \emph{et~al.}(2010)Jardine, Lee, Ward, Alexandrowicz,
  Hedgeland, Allison, Ellis, and Pollak]{jardine2010determination}
A.~Jardine, E.~Lee, D.~Ward, G.~Alexandrowicz, H.~Hedgeland, W.~Allison,
  J.~Ellis and E.~Pollak, \emph{Physical Review Letters}, 2010, \textbf{105},
  136101\relax
\mciteBstWouldAddEndPuncttrue
\mciteSetBstMidEndSepPunct{\mcitedefaultmidpunct}
{\mcitedefaultendpunct}{\mcitedefaultseppunct}\relax
\EndOfBibitem
\bibitem[Sanz \emph{et~al.}(2013)Sanz, Mart{\'\i}{\i}nez-Casado, and
  Miret-Art{\'e}s]{Ruth2013}
A.~Sanz, R.~Mart{\'\i}{\i}nez-Casado and S.~Miret-Art{\'e}s, \emph{Surface
  Science}, 2013, \textbf{617}, 229--232\relax
\mciteBstWouldAddEndPuncttrue
\mciteSetBstMidEndSepPunct{\mcitedefaultmidpunct}
{\mcitedefaultendpunct}{\mcitedefaultseppunct}\relax
\EndOfBibitem
\bibitem[Torres-Miyares and Miret-Artés(2025)]{Elena2025}
E.~E. Torres-Miyares and S.~Miret-Artés, \emph{Phys. Chem. Chem. Phys.}, 2025,
  \textbf{27}, 14739--14743\relax
\mciteBstWouldAddEndPuncttrue
\mciteSetBstMidEndSepPunct{\mcitedefaultmidpunct}
{\mcitedefaultendpunct}{\mcitedefaultseppunct}\relax
\EndOfBibitem
\bibitem[C.~T.~Chudley(1961)]{Chudley}
R.~J.~E. C.~T.~Chudley, \emph{Proceedings of the Physical Society}, 1961,
  \textbf{77}, 353\relax
\mciteBstWouldAddEndPuncttrue
\mciteSetBstMidEndSepPunct{\mcitedefaultmidpunct}
{\mcitedefaultendpunct}{\mcitedefaultseppunct}\relax
\EndOfBibitem
\bibitem[Torres-Miyares and Miret-Art{\'e}s(2025)]{Torres2025a}
E.~Torres-Miyares and S.~Miret-Art{\'e}s, \emph{Physical Chemistry Chemical
  Physics (submitted)}, 2025\relax
\mciteBstWouldAddEndPuncttrue
\mciteSetBstMidEndSepPunct{\mcitedefaultmidpunct}
{\mcitedefaultendpunct}{\mcitedefaultseppunct}\relax
\EndOfBibitem
\bibitem[Feynman \emph{et~al.}(1964)Feynman, Leighton, and
  Sands]{feynman1964brownian}
R.~Feynman, R.~Leighton and M.~Sands, \emph{The Feynman lectures of physics},
  1964, \textbf{1}, 41--1\relax
\mciteBstWouldAddEndPuncttrue
\mciteSetBstMidEndSepPunct{\mcitedefaultmidpunct}
{\mcitedefaultendpunct}{\mcitedefaultseppunct}\relax
\EndOfBibitem
\bibitem[Breuer and Petruccione(2002)]{breuer2002theory}
H.~Breuer and F.~Petruccione, \emph{``{T}he theory of open quantum systems''},
  Oxford University Press on Demand, 2002\relax
\mciteBstWouldAddEndPuncttrue
\mciteSetBstMidEndSepPunct{\mcitedefaultmidpunct}
{\mcitedefaultendpunct}{\mcitedefaultseppunct}\relax
\EndOfBibitem
\bibitem[Caldeira and A.J.(1983)]{Caldeira}
A.~Caldeira and L.~A.J., \emph{Annals of Physics}, 1983, \textbf{149},
  374\relax
\mciteBstWouldAddEndPuncttrue
\mciteSetBstMidEndSepPunct{\mcitedefaultmidpunct}
{\mcitedefaultendpunct}{\mcitedefaultseppunct}\relax
\EndOfBibitem
\bibitem[Schlosshauer(2007)]{schlosshauer2007decoherence}
M.~Schlosshauer, \emph{``{D}ecoherence: and the quantum-to-classical
  transition''d}, Springer Science \& Business Media, 2007\relax
\mciteBstWouldAddEndPuncttrue
\mciteSetBstMidEndSepPunct{\mcitedefaultmidpunct}
{\mcitedefaultendpunct}{\mcitedefaultseppunct}\relax
\EndOfBibitem
\bibitem[Mousavi and Miret-Art{\'e}s(2022)]{Mousavi-2022}
S.~Mousavi and S.~Miret-Art{\'e}s, \emph{The European Physical Journal Plus},
  2022, \textbf{137}, 1--14\relax
\mciteBstWouldAddEndPuncttrue
\mciteSetBstMidEndSepPunct{\mcitedefaultmidpunct}
{\mcitedefaultendpunct}{\mcitedefaultseppunct}\relax
\EndOfBibitem
\bibitem[Ford \emph{et~al.}(2001)Ford, Lewis, and
  O'Connell]{PhysRevA.64.032101}
G.~W. Ford, J.~T. Lewis and R.~F. O'Connell, \emph{Phys. Rev. A}, 2001,
  \textbf{64}, 032101\relax
\mciteBstWouldAddEndPuncttrue
\mciteSetBstMidEndSepPunct{\mcitedefaultmidpunct}
{\mcitedefaultendpunct}{\mcitedefaultseppunct}\relax
\EndOfBibitem
\bibitem[Ford and Connell(2003)]{Ford_2003}
G.~W. Ford and R.~F.~O. Connell, \emph{Journal of Optics B: Quantum and
  Semiclassical Optics}, 2003, \textbf{5}, S609--S612\relax
\mciteBstWouldAddEndPuncttrue
\mciteSetBstMidEndSepPunct{\mcitedefaultmidpunct}
{\mcitedefaultendpunct}{\mcitedefaultseppunct}\relax
\EndOfBibitem
\bibitem[Pe{\~{n}}ate-Rodr{\'{\i}}guez
  \emph{et~al.}(2012)Pe{\~{n}}ate-Rodr{\'{\i}}guez, Mart{\'{\i}}nez-Casado,
  Rojas-Lorenzo, Sanz, and Miret-Art{\'{e}}s]{Pe_ate_Rodr_guez_2012}
H.~C. Pe{\~{n}}ate-Rodr{\'{\i}}guez, R.~Mart{\'{\i}}nez-Casado,
  G.~Rojas-Lorenzo, A.~S. Sanz and S.~Miret-Art{\'{e}}s, \emph{Journal of
  Physics: Condensed Matter}, 2012, \textbf{24}, 104013\relax
\mciteBstWouldAddEndPuncttrue
\mciteSetBstMidEndSepPunct{\mcitedefaultmidpunct}
{\mcitedefaultendpunct}{\mcitedefaultseppunct}\relax
\EndOfBibitem
\bibitem[Chandrasekhar(1943)]{chandrasekhar1943stochastic}
S.~Chandrasekhar, \emph{Reviews of Modern Physics}, 1943, \textbf{15}, 1\relax
\mciteBstWouldAddEndPuncttrue
\mciteSetBstMidEndSepPunct{\mcitedefaultmidpunct}
{\mcitedefaultendpunct}{\mcitedefaultseppunct}\relax
\EndOfBibitem
\bibitem[Ward(2013)]{Ward2013}
D.~Ward, \emph{``{A} study of spin-echo lineshapes in helium atom scattering
  from adsorbates''}, 2013\relax
\mciteBstWouldAddEndPuncttrue
\mciteSetBstMidEndSepPunct{\mcitedefaultmidpunct}
{\mcitedefaultendpunct}{\mcitedefaultseppunct}\relax
\EndOfBibitem
\bibitem[Ellis \emph{et~al.}(1999)Ellis, Graham, and
  Toennies]{ellis1999quasielastic}
J.~Ellis, A.~Graham and J.~Toennies, \emph{Physical Review Letters}, 1999,
  \textbf{82}, 5072\relax
\mciteBstWouldAddEndPuncttrue
\mciteSetBstMidEndSepPunct{\mcitedefaultmidpunct}
{\mcitedefaultendpunct}{\mcitedefaultseppunct}\relax
\EndOfBibitem
\bibitem[Torres-Miyares \emph{et~al.}(2021)Torres-Miyares, Rojas-Lorenzo,
  Rubayo-Soneira, and Miret-Art{\'e}s]{torres2021surface}
E.~Torres-Miyares, G.~Rojas-Lorenzo, J.~Rubayo-Soneira and S.~Miret-Art{\'e}s,
  \emph{Mathematics}, 2021, \textbf{9}, 362\relax
\mciteBstWouldAddEndPuncttrue
\mciteSetBstMidEndSepPunct{\mcitedefaultmidpunct}
{\mcitedefaultendpunct}{\mcitedefaultseppunct}\relax
\EndOfBibitem
\bibitem[Vega \emph{et~al.}(2002)Vega, Guantes, and
  Miret-Art{\'e}s]{Guantes2002}
J.~Vega, R.~Guantes and S.~Miret-Art{\'e}s, \emph{Journal of Physics: Condensed
  Matter}, 2002, \textbf{14}, 6191\relax
\mciteBstWouldAddEndPuncttrue
\mciteSetBstMidEndSepPunct{\mcitedefaultmidpunct}
{\mcitedefaultendpunct}{\mcitedefaultseppunct}\relax
\EndOfBibitem
\bibitem[Vega \emph{et~al.}(2002)Vega, Guantes, and Miret-Artés]{Vega2002}
J.~L. Vega, R.~Guantes and S.~Miret-Artés, \emph{Phys. Chem. Chem. Phys.},
  2002, \textbf{4}, 4985--4991\relax
\mciteBstWouldAddEndPuncttrue
\mciteSetBstMidEndSepPunct{\mcitedefaultmidpunct}
{\mcitedefaultendpunct}{\mcitedefaultseppunct}\relax
\EndOfBibitem
\bibitem[Guantes \emph{et~al.}(2003)Guantes, Vega, Miret-Artés, and
  Pollak]{Guantes2003}
R.~Guantes, J.~L. Vega, S.~Miret-Artés and E.~Pollak, \emph{The Journal of
  Chemical Physics}, 2003, \textbf{119}, 2780--2791\relax
\mciteBstWouldAddEndPuncttrue
\mciteSetBstMidEndSepPunct{\mcitedefaultmidpunct}
{\mcitedefaultendpunct}{\mcitedefaultseppunct}\relax
\EndOfBibitem
\bibitem[Gradshteyn and Ryzhik(2014)]{Table}
I.~S. Gradshteyn and I.~M. Ryzhik, \emph{Table of integrals, series, and
  products}, Academic press, 2014\relax
\mciteBstWouldAddEndPuncttrue
\mciteSetBstMidEndSepPunct{\mcitedefaultmidpunct}
{\mcitedefaultendpunct}{\mcitedefaultseppunct}\relax
\EndOfBibitem
\end{mcitethebibliography}

\end{document}